\documentstyle[sprocl]{article}
\newcommand{\be}{\begin{eqnarray}}
\newcommand{\ee}{\end{eqnarray}}

\def\be{\begin{equation}}
\def\ee{\end{equation}}
\def\bea{\begin{eqnarray}}
\def\eea{\end{eqnarray}}

\begin{document}

\title{HOW TO FIND THE QCD CRITICAL POINT\footnote{Work 
done in collaboration with Misha Stephanov and 
Edward Shuryak.\cite{us1}}}

\author{Krishna Rajagopal}

\address{Center for Theoretical Physics, MIT, Cambridge, MA 02139, USA\\
E-mail: krishna@ctp.mit.edu\\~~ \\{\rm March 30, 1999. MIT-CTP-2848}}


\maketitle\abstracts{The event-by-event fluctuations in
heavy ion collisions carry information about the thermodynamic
properties of the hadronic system at the time of freeze-out. By studying
these fluctuations as a function of varying control parameters,
such as the collision energy, it is possible to learn much
about the phase diagram of QCD. As a timely example, we stress
the methods by which present experiments at the CERN SPS can
locate the second order critical point at which a line of first
order phase transitions ends. Those event-by-event signatures
which are characteristic of freeze-out in the vicinity of the 
critical point will exhibit nonmonotonic dependence
on control parameters. We focus on observables constructed from 
the multiplicity and transverse momenta of charged
pions. We find good agreement between NA49 data and 
thermodynamic predictions for the noncritical
fluctuations of such observables. We then analyze 
the effects due to the critical fluctuations of the sigma field.
We estimate the size of these nonmonotonic effects which 
appear near the critical point, including restrictions
imposed by finite size and finite time.}

\section{Introduction}

In my talk at Strong and Electroweak Matter '98, I presented
recent work on the physics which 
arises in two different areas of the 
QCD phase diagram.  Cold dense 
quark matter forms a color superconductor, and I compared
the superconducting phase expected in QCD with two massless
quarks with that expected for 
three massless quarks in which
chiral symmetry is broken by color-flavor locking.
Alford, Berges and I have recently 
completed an
analysis of 
the phase diagram of zero temperature QCD as a function
of density and strange quark mass, in Ref. \cite{ABR}. This
brings the ideas I presented in Copenhagen together into 
a consistent picture, and
I refer you to that paper for an up-to-date treatment
of the subject, and references to the literature.

The other
half of my talk in Copenhagen was a sketch of methods by
which present heavy ion experiments can find the critical point
in the QCD phase diagram at nonzero temperature $T$ 
and baryon chemical potential $\mu$.  Like the end point
in the electroweak phase diagram, discussed by others
at this meeting, this critical point is a second order transition
in the Ising universality
class which occurs at the end of a line of first order phase
transitions.  Stephanov, Shuryak and I have recently completed
a detailed analysis of the signatures of the physics
characteristic of the vicinity of this point,\cite{us1} 
begun in Ref. \cite{SRS}. I described this work in a preliminary
fashion in Copenhagen; in these proceedings,
I summarize the results and implications
of Ref. \cite{us1}. Those interested
in the derivation of these results should see Ref. \cite{us1}.

Large acceptance detectors, such as NA49 and WA98 at CERN,
have made it possible to measure important average
quantities in single heavy ion collision events.
For example, instead of analyzing the
distribution of charged particle
transverse momenta obtained by averaging over particles
from many events, we can now study the event-by-event
variation of the mean transverse momentum 
of the charged pions in a single event, $p_T$.\footnote{We 
denote the mean transverse momentum 
of all the pions in a single event by $p_T$ rather than 
$\langle p_T\rangle$ because we choose to reserve $\langle \ldots \rangle$
for averaging over an ensemble of events.}
Although much of this data still has preliminary status, with
more statistics and more detailed analysis yet to come, some general
features have already been
demonstrated. In particular, the event-by-event 
distributions of these observables are as perfect 
Gaussians as the data statistics allow, and the
fluctuations --- the width
of the Gaussians --- are small.\cite{trento}

This is very different from what one observes in $pp$ collisions, 
in which fluctuations are large.
These large non-Gaussian fluctuations clearly
reflect non-trivial quantum
fluctuations, all the way from the nucleon wave function to that of the
secondary hadrons, and are not yet sufficiently well understood.
As discussed in Refs. \cite{GM92,GLR99}, thermal
equilibration in $AA$ collisions drives the variance of the event-by-event
fluctuations down, close to the value determined by the variance of 
the inclusive one-particle
distribution divided by the square root of the multiplicity.

Can we learn something from
the magnitude of these small fluctuations and their dependence on the
parameters of the collision?  What do the widths of
the Gaussians tell us about the thermodynamics of QCD?
Some of these questions have been addressed in 
Refs. \cite{Stodolsky,Shu_fluct} where it was pointed out that,   
for example, temperature fluctuations are related to heat capacity via
\be \label{cv}
{\langle (\Delta T)^2 \rangle\over T^2} = {1 \over C_V(T) },
\ee
and so can tell us about thermodynamic properties of
the matter at freeze-out. 
Furthermore, Mr\'owczy\'nski has discussed the study of the
compressibility of hadronic matter at freeze-out via the
event-by-event fluctuations of the particle number \cite{Mrow1} 
and Ga\'zdzicki \cite{Gaz} and
Mr\'owczy\'nski \cite{Mrow2} have considered event-by-event
fluctuations of the kaon to pion ratio as measured
by NA49 \cite{trento}.  
In $pp$ physics one can hope to extract 
quantum mechanical information about the initial state
from event-by-event fluctuations of the final state;
in heavy ion collisions 
equilibration 
renders this an impossible goal. 
In $AA$ collisons, then, the new goal is
to use the much smaller, Gaussian
event-by-event fluctuations of the final state
to learn about thermodynamic properties at freeze-out. 

It is worth noting that once a large
acceptance detector has presented convincing evidence that the 
event-by-event distribution of, for example, $p_T$ is Gaussian,
then the measurement of the width of such a distribution can
be accomplished by ``event-by-event'' measurements in which
only two pions per event are observed. This
has recently been emphasized by Bia\l as and Koch.\cite{Bialas}  Of course,
this approach measures the width of the event-by-event distribution 
whether or not it is Gaussian; it is only the results of a
large acceptance experiment like NA49 which motivate
a thermodynamic analysis of the event-by-event fluctuations.

Stephanov, Shuryak and I focus on observables constructed
from the multiplicity and the momenta of the charged particles in the final
state, as measured by NA49.  
We leave the extension of
the methods of this paper to the study of thermodynamic
implications of the NA49
Gaussian distribution of event-by-event $K/\pi$ ratios \cite{trento}
and of the WA98 Gaussian distribution of event-by-event
$\pi^0/\pi^\pm$ ratios \cite{steinberg} for future work.

One of the lessons of our paper is that it is difficult
to apply thermodynamic relations like (\ref{cv}) directly.
To see a sign of this, note that 
the event-by-event
fluctuations of the energy $E$ of a part of a finite 
system in thermal equilibrium are given by
$\langle (\Delta E)^2\rangle = T^2 C_V(T)$.
For a system in equilibrium, the mean values of $T$ and $E$ 
are directly related by an equation of state $E(T)$;
their fluctuations, however,
have quite different behavior as a function of $C_V$, and
therefore behave differently when $C_V$ diverges at a critical point.
The fluctuations of ``mechanical'' observables increase
at the critical point. Because $T(E)$ is singular at
the critical point, the fluctuations of $T$ decrease there.
It is a fact that what we measure are the mechanical observables, 
and since we in general only know $T(E)$ for simple systems
we call thermometers, we cannot apply (\ref{cv}) to
the complicated system of interest.  It is not in fact
necessary to translate the observed ``mechanical'' variable
(the mean transverse momentum $p_T$ for example) into a temperature in order to
detect the critical point.  It is easier to look directly
at the fluctuations of observable quantities. We demonstrate
that 
the fluctuations of $p_T$ grow at the critical point. 

Although our methods are general, we focus in Ref. \cite{us1}
on how to use them to
find and study the critical end-point
E on the phase diagram of QCD in the $T\mu$ plane.  The possible
existence of such a point, as an endpoint of the first order
transition separating quark-gluon plasma from hadron matter, and its
universal critical properties have been pointed out recently 
in Refs. \cite{BeRa97,HaJa97}. The point E can be thought of as a
descendant of a tricritical point in the phase diagram for 
2-flavor QCD with
{\em massless} quarks.  In a previous letter, we have laid
out the basic ideas for observing the critical endpoint~\cite{SRS}.
The signatures 
proposed in Ref. \cite{SRS} are based on the fact that such a point is a
genuine thermodynamic singularity at which 
susceptibilities diverge and the order parameter
fluctuates on long wavelengths. The resulting
signatures all share one common property: they are {\em
nonmonotonic\/} as a function of an experimentally varied parameter
such as the collision energy, centrality, rapidity or 
ion size. 
Once experimentalists vary a control parameter
which causes 
the freeze-out point in the $(T,\mu)$ plane to move toward, through, 
and then past the vicinity of the endpoint E, they 
should see all the signatures we describe first strengthen, 
reach a maximum,  
and then decrease, as a nonmonotonic function
of the control parameter.  It is important to have a control
parameter whose variation changes the $\mu$ at which
the system crosses the transition region and freezes out.
The collision energy is an obvious choice, since it is
known experimentally that varying the collision energy
has a large effect on $\mu$ at freeze-out.  Other possibilities
should also be explored.\footnote{If the system crosses the
transition region near E, but only freezes out at a much
lower temperature, the event-by-event fluctuations will not
reflect the thermodynamics near E.  In this case, one can
push freeze-out to earlier times and thus closer to E by
using smaller ions.\cite{SRS}}

We assume throughout 
that freeze-out occurs from an equilibrated hadronic system.  
If freeze-out occurs ``to the left'' (lower $\mu$;
higher collision energy) of the critical end point E, it occurs
after the matter has traversed the crossover region
in the phase diagram. If it occurs
``to the right'' of E, it occurs after the matter has traversed
the first order phase transition.  This is the situation in 
which our assumption of freeze-out from an equilibrated system
is most open to question. First, one may imagine hadronization
directly from the mixed phase, without time for the hadrons
to rescatter.  Hadronic elastic scattering cross-sections are
large enough that this is unlikely.  Second, one may worry
that the matter is inhomogeneous after the first order
transition, and has not had time to re-equilibrate.
Fortunately, our assumption is testable.
If the matter were inhomogeneous at freeze-out, 
one can expect non-Gaussian fluctuations
in various observables \cite{heiselberg} which
would be seen in the same experiments that seek 
the signatures we describe.  We focus
on the Gaussian thermal fluctuations of an equilibrated
system, and study the nonmonotonic changes in these
fluctuations associated with moving the freeze-out point toward and then
past the critical point, for example from left to right as
the collision energy is reduced.



Ref. \cite{us1} is devoted to a detailed
analysis of the physics behind event-by-event fluctuations in
relativistic heavy ion collisions and 
the resulting effects unique to the vicinity 
of the critical point in the phase 
diagram of QCD.
Most of our analysis is applied to the fluctuations
of the observables characterizing the multiplicity and 
momenta of the charged pions in the final state of a
heavy ion collision. 
There are several reasons why the pion observables
are most sensitive to the critical fluctuations. First, the
pions are the most numerous hadrons produced and observed 
in relativistic heavy ion collisions.
A second, very important reason, is that pions couple strongly to 
the fluctuations of the sigma field (the magnitude of
the chiral condensate) which is
the order parameter of the phase transition.  Indeed, the pions
are the quantized oscillations of the phase of the chiral
condensate and so it is not surprising that at the critical
end point, where the magnitude of the condensate is fluctuating
wildly, signatures are imprinted on the pions.  

\section{Noncritical Thermal Fluctuations in Heavy Ion Collisions}

Before we discuss the effects of the critical fluctuations,
we must analyze the thermal fluctuations which are present
if freeze-out does {\it not} occur in the vicinity of the
critical point.  In this section, but not
throughout this paper, we assume that the system freezes
out far from the critical point in the phase diagram, and
can be approximated as an ideal resonance gas when it
freezes out.  
We compare some of our results to preliminary data from 
the NA49 experiment
on PbPb collisions at 160 AGeV, and find broad agreement.
The results obtained seem to support the hypotheses 
that 
most 
of the fluctuation observed in the data
is indeed thermodynamic in origin, and that this system
is not freezing out near the critical point.  

As a first test of our resonance gas model,
we analyze the fluctuations
in an ideal Bose gas of pions, and then add as many 
of the effects which this simple treatment neglects
except that we assume that no effects due to critical
fluctuations are significant. 
We model the matter in a relativistic
heavy ion collision at freeze-out as
a resonance gas in thermal equilibrium,
and begin by calculating
the variance of the event-by-event 
fluctuations of total multiplicity $N$. 
The fluctuations in $N$ are not affected by
the boost which the pion momenta receive from the collective flow,
but they are contaminated 
experimentally by
fluctuations in the impact parameter.  
This experimental
contamination can be reduced by 
making a tight enough centrality cut 
using a zero degree calorimeter.

We find $\langle (\Delta N)^2\rangle/\langle N \rangle\approx 1.5$,
which we compare with NA49 results from central Pb-Pb collisions
at 160 AGeV.
It is clear that with no cut on centrality, one would see 
a very wide non-Gaussian
distribution of multiplicity determined by
the geometric probability of different impact parameters $b$.
Gaussian thermodynamic fluctuations can only be seen if
a tight enough cut in centrality is applied.  
The event-by-event $N$-distribution found by NA49 when
they use only the $5\%$ most central of all events,
with centrality measured using a zero degree calorimeter,
is Gaussian to within about $5\%$.  This cut corresponds
to keeping collisions with impact parameters $b<3.5$ fm.\cite{trento}
The non-Gaussianity
could be further reduced by tightening the centrality
cut further. 
From the data, we have $\langle (\Delta N)^2\rangle/\langle N \rangle
= 2.008\pm 0.009$, which suggests that about $75\%$ of the
observed
fluctuation is thermodynamic in origin.  The 
contamination introduced into the data by fluctuations in centrality 
could be reduced by
analyzing
data samples with more or less restrictive
cuts but the same $\langle N\rangle$, and extrapolating to a limit
in which the cut is extremely restrictive.
This could be done using
cuts centered at any centrality.  
Our resonance gas model predicts that
as the centrality cut is tightened, 
the ratio $v^2_{\rm ebe}(N)/\langle N \rangle$ should decrease toward 
a limit near 1.5.

Although further work is certainly required, it is already apparent
that the bulk of the multiplicity fluctuations observed
in the data are thermodynamic in origin.  
impact parameter. 
Note that our prediction is strongly
dependent on the presence of the resonances; had
we not included them, our prediction would have been
significantly lower, farther below the data.
Because the multiplicity
fluctuations are sensitive to impact parameter fluctuations,
it may prove difficult to  explain  
their magnitude with greater precision
even in future. However, the fact that they are largely thermodynamic
in origin
suggests that the effects present near the critical point, which
we describe below, could result
in a
significant nonmonotonic enhancement of the multiplicity fluctuations.  
This would be of interest
whether or not the noncritical fluctuations on top of which
the nonmonotonic variation 
occurs are understood with precision. 

We then turn to a 
calculation of the variance of the event-by-event
fluctuations of the mean transverse momentum, $p_T$.
We first calculate the width of the {\it inclusive}
$p_T$-distribution, $v_{\rm inc}(p_T)$.  In the absence
of any correlations, the event-by-event fluctuations of
the mean transverse momentum of the charged pions in an event,
$v_{\rm ebe}(p_T)\equiv\langle (\Delta p_T)^2\rangle^{1/2}$ 
would be given by $v_{\rm inc}(p_T)/\langle N \rangle^{1/2}$,
and this turns out to be a very good approximation in the present
data as we discuss below.
We calculate numerically the contribution to $v_{\rm inc}(p_T)$
from ``direct pions'',
already present at freeze-out, and from the pions generated later
by resonance decay.  We have simulated a gas of pions, nucleons and resonances
in thermal equilibrium at freeze-out, including the 
$\pi$, $K$, $\eta$, $\rho$, $\omega$, $\eta'$, $N$, $\Delta$,
$\Lambda$, $\Sigma$ and $\Xi$,
and then simulated
the subsequent decay of the resonances. 
That is, we have generated an ensemble of pions 
in three steps:  (i) Thermal ratios of hadron multiplicities
were calculated  assuming equilibrium ratios at chemical
freeze-out. Following \cite{PBM_etal}, the values 
$T_{\rm ch}=170$ MeV and
$\mu_{\rm baryon}=200$ MeV  were used. (ii)  
Then, a program  generates 
hadrons with multiplicities determined at chemical
freezeout, but with thermal momenta as appropriate  
at the thermal freeze-out  temperature, 
which we take to be $T_{\rm f}=120$ MeV, with
$\mu_\pi=60$ MeV.  The last step (iii) is to
decay all the resonances. From the resulting
ensemble of pions (the sum of the direct pions and those from the 
resonances) we obtain $v_{\rm inc}(p_T)/\langle p_T\rangle = 0.66$
The resonances
turn out to be less important here than in the 
calculation of the multiplicity fluctuations, in that the
resonance gas prediction for 
$v_{\rm inc}(p_T)/\langle p_T\rangle $
is almost indistinguishable from that of an 
ideal Bose gas of pions.

To this point, we have calculated the fluctuations
in $p_T$ as if the matter in a heavy ion
collision were at rest at freeze-out.  This is
not the case: by that stage  
the hadronic matter is undergoing a collective hydrodynamic
expansion in the transverse direction, and this
must be taken into account in order to compare
our results with the data.
A very important point here is that the fluctuations
in pion multiplicity  
are not affected by flow, and our prediction for them
is therefore unmodified.
However the event-by-event 
fluctuations of mean $p_T$ are certainly affected by flow.
The fluctuations we have calculated pertain to the rest
frame of the matter at freeze-out, and we must now boost
them. A detailed account of the resulting effects would
require a complicated analysis.  We use 
the simple approximation \cite{SSH} that the effects of
flow on the pion momenta can be treated as a Doppler
blue shift of the spectrum: $n(p_T)\to
n(p_T\sqrt{1-\beta}/\sqrt{1+\beta})$.  This blue shift
increases $\langle p_T\rangle$, and increases $v_{\rm inc}(p_T)$,
but leaves the ratio $v_{\rm inc}(p_T)/\langle p_T\rangle$
(and therefore the ratio $v_{\rm ebe}(p_T)/\langle p_T\rangle$)
unaffected. 
However, event-by-event fluctuations in
the flow velocity $\beta$ 
must still be taken into account. 
This issue was 
discussed qualitatively already in \cite{Shu_fluct}, where it was
argued that this effect must be relatively weak. In Ref. \cite{us1}
we provide the first rough estimate of its 
magnitude. We estimate  that fluctuations in the flow velocity increase
$v_{\rm inc}(p_T)/\langle p_T\rangle$
from $0.66$ to $0.67$.
The largest uncertainty in our estimate for 
$v_{\rm inc}(p_T)/\langle p_T\rangle$ is not due to the fluctuations
in the flow velocity, which can clearly be neglected, but
is due to the velocity itself.  The blue shift approximation
which we have used applies quantitatively only to 
pions with momenta greater than their mass \cite{SSH}.
Because of the nonzero pion mass, boosting the pions
does not actually scale the momentum spectrum by 
a momentum independent factor. Furthermore, in a real
heavy ion collision there will be a position dependent
profile of velocities, rather than a single velocity $\beta$.
A more complete calculation of $v_{\rm inc}(p_T)/\langle p_T\rangle$
would require a better treatment of these effects in a hydrodynamic
model; we leave this for the future.

We compare our results to the NA49 data, in which 
$v_{\rm inc}(p_T)/\langle p_T\rangle = 0.749\pm 0.001$.
We see that the major part of the observed fluctuation in $p_T$ 
is accounted for by the thermodynamic fluctuations
we have considered.  
Our prediction 
is about $10\%$ lower than that in the 
data.
First, this suggests
that there may be a small nonthermodynamic contribution
to the $p_T$-fluctuations, for example from fluctuations
in the impact parameter.  (However, we expect that the 
fluctuations of an intensive quantity like $p_T$ are less
sensitive to impact parameter fluctuations than are those
of the multiplicity, and this seems to be borne out by the data.)
The other source of the discrepancy is the blue shift approximation.
We leave a more sophisticated treatment
of the effects of flow on the spectrum to
future work. Such a treatment is necessary before 
we can estimate how much of the 
$10\%$ discrepancy is introduced by 
the blue
shift approximation.  Future work on the experimental
side (varying the centrality cut) could lead to
an estimate of how much of the discrepancy is due
to impact parameter fluctuations.

We have gone as far as we will go in this paper 
in our quest to understand
the thermodynamic origins of the width of the inclusive
single particle distribution. 
We now turn 
to the ratio 
of the scaled  event-by-event variation to the variance of the 
inclusive distribution:
\begin{equation}
\sqrt{F} \equiv 
{\langle N\rangle^{1/2} v_{\rm ebe}(p_T)\over v_{\rm inc}(p_T)} 
= 1.002 \pm 0.002.
\label{dataratio}
\end{equation}
The difference between the scaled event-by-event variance
and the variance of the inclusive distribution is 
less than a percent in the NA49 data.\footnote{We 
explain in an Appendix in Ref. \cite{us1} that
in order to be sure that $F=1$ when there are no correlations
between pions, care must be taken in constructing an 
estimator for $v_{\rm ebe}(p_T)$ using a finite sample
of events, each of which has finite multiplicity. 
The appropriate 
prescription is to
weight events in the event-by-event average by their
multiplicity, and we have made the appropriate
correction in writing (\ref{dataratio}).  Other authors \cite{GM92}
have introduced the correlation measure 
$\Phi_{p_T}= \langle N \rangle^{1/2} v_{\rm ebe}(p_T) - v_{\rm inc}(p_T)$.
Because $v_{\rm inc}(p_T)$ is scaled by the blue shift
introduced by the expansion velocity, so is 
$\Phi_{p_T}$. This makes $\Phi_{p_T}$ harder 
to predict than $F$.  However, for convenience, we note that if
one uses the experimental value of $v_{\rm inc}(p_T)$, a
value $\sqrt{F}=1.01$ corresponds to $\Phi_{p_T}=2.82$ MeV,
and the $\sqrt{F}$ in the data (\ref{dataratio}) corresponds
to $\Phi_{p_T}=0.6\pm 0.6$ MeV.}

We analyze a number of noncritical
contributions to the ratio $\sqrt{F}$,
which we write 
\begin{equation}
\sqrt{F}=\sqrt{F_B F_{\rm res} F_{\rm EC} } \ .
\label{theoryratio}
\end{equation}
$F_B$ is the contribution of the Bose enhancement of the
fluctuations of identical pions. We calculate this effect and find
$\sqrt{F_B} \approx 1.02$.
$F_{\rm res}$ describes the effect of the correlations induced by
the fact that pions produced by the decay of a resonance
after freeze-out do not have a chance to rescatter.
We estimate it by dividing the pions from our 
resonance gas simulation into ``events'' of varying
sizes, and evaluating $F$. Since Bose enhancement
is not included in the simulation, the $F$ so obtained
is just $F_{\rm res}$. We find no statistically significant
contribution, and conclude that $|F_{\rm res}-1|<0.01$.

The third contribution, $F_{\rm EC}$, is due to energy conservation
in a finite system.  This is most easily described 
by considering the event-by-event fluctuations $\Delta n_p$ 
in the number of pions in a bin in momentum space centered at
momentum $p$.  Consider the correlator $\langle \Delta n_p\Delta n_k \rangle$.
When one $n_p$ fluctuates up, others must
fluctuate down, and it is therefore more likely
that $n_k$ fluctuates downward.  Energy conservation in
a finite system therefore leads to an anti-correlation
which is off-diagonal in $pk$ space. $v_{\rm ebe}(p_T)$ is determined
by $\langle \Delta n_p\Delta n_k \rangle$, and the result
of this anti-correlation is a reduction:
\begin{equation}
\sqrt{F_{\rm EC}}\approx 0.99\ .
\label{FECest}
\end{equation}
If the observed charged pions are in thermal contact with an
unobserved heat bath, 
the anti-correlation introduced by energy conservation
decreases as the
heat capacity of the heat bath increases.   
The estimate (\ref{FECest}) assumes that the heat capacity
of the direct charged pions is about $1/4$ of the total
heat capacity of the hadronic system at freezeout.
In addition to the contributions we calculate, $\sqrt{F}$
is affected by the finite two-track resolution in the
detector, and by final state Coulomb interactions between
charged pions.  
NA49 estimates that these contributions reduce $F$ by about
the same amount that Bose enhancement increases it.

We conclude that the ratio $\sqrt{F}$ measured by NA49
is broadly consistent with thermodynamic expectations.
It receives a positive contribution from Bose enhancement,
negative contributions from energy conservation and
two-track resolution, and a positive contribution
from the effect of resonance decays.  These contributions
to $\sqrt{F}$ are all roughly at the $1\%$ level (or 
smaller in the case of that from resonance decays)
and it seems that they cancel in the data (\ref{dataratio}).
Our results support the
general idea that the small fluctuations
observed in $AA$ collisions, relative to those in $pp$,
are consistent with the hypothesis that the matter in
the $AA$ collisions achieves approximate local thermal
equilibrium in the form of a resonance gas. 

With more detailed experimental study, either now
at the SPS, or soon at RHIC (STAR will study 
event-by-event fluctuations in $p_T$, $N$, particle ratios, etc;
PHENIX and PHOBOS in $N$ only)
it should be possible 
to disentangle the different effects we describe.
Making a cut to look at only low $p_T$ pions should increase the effects of 
Bose enhancement. 
The anti-correlation
introduced by energy conservation 
is due
to terms in $\langle \Delta n_p\Delta n_k \rangle$
which are off-diagonal in $pk$. Thus, a direct measurement
of $\langle \Delta n_p\Delta n_k \rangle$ would make it
easy to separate this anti-correlation from other effects.
The cross correlation $\langle \Delta N \Delta p_T\rangle$
is also a very interesting observable to study. It vanishes
for a classical ideal gas. This means that whereas $v_{\rm ebe}(p_T)$
receives a dominant contribution from the width of
the inclusive single particle distribution, this effect
cancels in $\langle \Delta N \Delta p_T\rangle$ and the
remaining effects due to Bose enhancement and energy conservation
dominate.  Although this cross-correlation is small, it is
worth measuring because it only receives contributions
from interesting effects.

We hope that the combination of the theoretical tools we
have provided and the present NA49 data provide a solid
foundation for the future study of the thermodynamics of
the hadronic matter present at freeze-out in heavy ion collisions. 
Once data is available for other collision energies, centralities
or ion sizes, the present NA49 data and the calculations
of this section will provide an experimental and
a theoretical baseline for the study of variation as
a function of control parameters.

Our analysis demonstrates
that the observed fluctuations
are broadly consistent with thermodynamic expectations, and 
therefore raises the possibility of large effects when
control parameters are changed in such a way that 
thermodynamic properties are changed significantly,
as at a critical point.  The smallness of the 
statistical errors in the data also highlights the possibility
that many of the interesting systematic effects we analyze
in this paper will be accessible to detailed study as
control parameters are varied.

\section{Pions Near the Critical Point: Interaction with the Sigma Field}

With the foundations established, we now describe
how the fluctuations we analyze will change if control
parameters are varied in
such a way that the baryon chemical potential  
at freeze-out, $\mu_{\rm f}$, moves toward and then past the critical
point in the QCD phase diagram at which a line 
of first order transitions ends at a second order endpoint.
The good agreement
between the noncritical thermodynamic fluctuations
we analyze in Section 2 and NA49 data make it unlikely
that central PbPb collisions at 160 AGeV freeze out
near the critical point.
Estimates we have made in Ref. \cite{SRS}
suggest that the critical point is
located at a baryon chemical potential 
$\mu$ such that it will be found at
an energy between 160 AGeV and AGS energies. This makes it
a prime target for detailed study at the CERN SPS
by comparing data taken at 40 AGeV, 160 AGeV, and in between.
If the critical
point is located at such a low $\mu$ that the maximum
SPS energy is insufficient to reach it, 
it would
then be in a regime accessible to study by the 
RHIC experiments.  We want to stress that
we are more confident
in our ability to describe the properties of the
critical point and thus to predict {\it how} to find it than
we are in our ability to predict where it is.

We now describe how the fluctuations of the pions will be
affected if the system freezes out near the critical
endpoint. First, because the pions at freeze-out are
now in contact with a heat bath whose heat capacity
diverges at the critical point, the effects of
energy conservation parametrized by $F_{\rm EC}-1$ are
greatly reduced. However, since $F_{\rm EC}$ is close
to one even away from the critical point, this is
a small effect.

The dominant effects of the critical 
fluctuations on the pions are the direct effects occuring
via the $\sigma \pi\pi$ coupling.  
In the previous section, we made the assumption that the ``direct
pions'' at freeze-out could be described as an ideal Bose gas.
We do not expect this to be a good approximation if
the freeze-out point is near the critical point.
The sigma field is the order parameter for the
transition and near the critical point it therefore develops large critical
long wavelength fluctuations. 
These fluctuations are responsible for singularities
in thermodynamic quantities.  
We find that because of the $G\sigma\pi\pi$ coupling,
the fluctuations
of both the multiplicity and 
the mean transverse momentum of the charged pions do in fact diverge
at the critical point.

We then estimate the size of the effects in a heavy
ion collision. This requires first estimating the strength
of the coupling constant $G$, and then taking into account
the finite size of the system and the finite time during
which the long wavelength fluctuations
can develop.
We find a large increase in the fluctuations
of both the multiplicity and 
the mean transverse momentum of the pions. This increase
would be divergent in the infinite volume limit 
precisely at the
critical point. We apply finite size and finite time
scaling to estimate how close the system created in
a heavy ion collision can come to the critical singularity,
and consequently how large an effect can be seen
in the event-by-event fluctuations of the pions.  We conclude that
the nonmonotonic changes in the variance of the event-by-event
fluctuation of the pion multiplicity and momenta
which are induced
by the universal physics characterizing the critical point
can easily be between one and two orders of magnitude
greater than the statistical errors in the
present data.

The value of the coupling $G$ in vacuum can be estimated either from the
relationship between the sigma and pion masses and $f_\pi$ or from 
the width of the sigma.  Both yield an 
estimate $G\sim 1900$ MeV, where we have used $m_\sigma=600$ MeV.  
The width of the sigma is so large that this ``particle''
is only seen as a broad bump in the $s$-wave $\pi-\pi$
scattering cross-section. The vacuum $\sigma \pi \pi$ 
coupling must be at least as large as $G\sim 1900$ MeV,
since the sigma would otherwise be too narrow.

The vacuum value of $G$ 
would not change much if one were to take the chiral limit
$m\rightarrow 0$. The situation is different at the critical
point.  Taking the quark mass to zero while following
the critical endpoint leads one to the tricritical point P
in the phase diagram for QCD with two massless quarks.
At this point, $G$ vanishes as we discuss below.
This suggests that at E, the coupling $G$ is less than in vacuum.
In Ref. \cite{us1}, we
use what we know about physics near the tricritical point P
to make an estimate of how much the coupling $G$ is 
reduced at the critical endpoint E (with the quark mass
$m$ having its physical value), relative to the vacuum value 
$G\sim 1900$ MeV
estimated above.

We begin by recalling some known results. (For details,
see Refs. \cite{BeRa97,HaJa97,SRS}.)
In QCD with two massless quarks, a spontaneously broken
chiral symmetry is restored at finite temperature. This
transition is likely second order and belongs in the universality
class of $O(4)$ magnets in three dimensions.  At zero $T$,
various models suggest that the chiral symmetry restoration
transition at finite $\mu$ is first order. Assuming that
this is the case, one can easily argue that there 
must be a tricritical point P in the $T\mu$ phase
diagram, where the transition changes from first
order (at higher $\mu$ than P) 
to second order (at lower $\mu$), and such a tricritical
point has been found in a variety of models.\cite{BeRa97,HaJa97,italians}
The nature of this point can be understood by considering
the Landau-Ginzburg effective potential for $\phi_\alpha$,
order parameter of chiral symmetry breaking:
\begin{equation}\label{phi6potential}
\Omega(\phi_\alpha) = \frac a2 \phi_\alpha\phi_\alpha 
+ \frac b4 (\phi_\alpha\phi_\alpha)^2 + \frac c6 (\phi_\alpha\phi_\alpha)^3\ .
\end{equation}
The coefficients $a$, $b$ and $c>0$ are functions of $\mu$
and $T$. The second order phase transition line described by
$a=0$ at $b>0$ becomes first order when $b$ changes sign, 
and the tricritical point P is therefore the point at which $a=b=0$.
The critical properties of this point can be inferred from
universality \cite{BeRa97,HaJa97}, and the exponents are as
in the mean field theory (\ref{phi6potential}). We will
use this below.  Most important in the present
context is the fact that
because $\langle \phi\rangle=0$ at P, there is 
no $\sigma\pi\pi$ coupling, and $G=0$ there.

In real QCD with nonzero quark masses, the second order phase transition
becomes a smooth crossover and the tricritical point P becomes E, the 
second order critical endpoint of a first order phase
transition line. Whereas at P there are four massless 
scalar fields undergoing critical long wavelength fluctuations,
the $\sigma$ is the only field which becomes massless
at the point E, and the point E is therefore in the Ising
universality class \cite{BeRa97,HaJa97}. The pions remain
massive at E because of the explicit chiral symmetry
breaking introduced by the quark mass $m$.  Thus, 
when we discuss physics near E as a function of $\mu$ and
$T$, but at fixed $m$, we will use universal scaling relations
with exponents from the three dimensional Ising model.  
Our present purpose, however, is to imagine varying $m$
while changing $T$ and $\mu$ in such a way as to stay
at the critical point E, and ask how large $G$ (and $m_\pi$)
become once $m$ is increased from zero (the tricritical
point P at which $G=m_\pi=0$) to its physical value.
For this task, we use exponents describing universal
physics near P.  Applying tricritical scaling relations
all the way up to a quark mass which is large
enough that $m_\pi$ is not small compared to $T_c$ may
introduce some uncertainty into our estimate.

We first determine the trajectory of 
the critical line
of Ising critical points E as a function of quark 
mass $m$,\footnote{See Ref. \cite{rajreview} for a derivation
of the analogous line of Ising points emerging from the tricritical
point in the QCD phase diagram at zero $\mu$ as a function of
$m$ and the strange quark mass $m_s$. This tricritical point can
be related to the one we are discussing by varying $m_s$.\cite{SRS}}
and then find that $G\sim m^{3/5}$ along this line, where
$m$ is the light quark mass. 
Thus the coupling $G$ is suppressed
compared to its ``natural'' vacuum value $G_{\rm vac}$ by
a factor of order $(m/\Lambda_{\rm QCD})^{3/5}$.
Taking $\Lambda_{\rm QCD}\sim 200$ MeV, $m\sim 10$ MeV we
obtain our estimate 
\begin{equation}
G_E\sim \frac{G_{\rm vac}}{6} \sim 300 {\rm ~MeV}\ .
\end{equation}
The main 
source of uncertainty in this estimate 
is our inability
to compute the various nonuniversal
masses which enter the estimate as prefactors in front
of the $m$ dependence which we have followed. In other words,
we do not know the correct value to use for 
$\Lambda_{\rm QCD}$ in the suppression factor 
which we write as $(m/\Lambda_{\rm QCD})^{3/5}$.

The final ingredient we need is an estimate of the 
correlation length $\xi$ of the sigma field, which
is infinite at the critical point.  In practice,
there are important restrictions on how large $\xi$
can become.  Two particle interferometry \cite{HeinzJacak}
suggests that the size of regions over which
freeze-out is homogeneous is roughly $12$ fm in both
the longitudinal and transverse directions. This means
that the finite size of the system limits $\xi$ to be less
than about this value. The finite time restriction
is stricter, but harder to estimate.\cite{HoHa} Although the 
size of the system allows the correlation length
to grow to 12 fm, there may not be enough time
for such long correlations to grow.  We use 
$\xi_{\rm max}\sim 6$ fm as a rough estimate of the largest
correlation length possible if control parameters
are chosen in such a way that the system freezes out
close to the critical point.

We now return to our discussion of the effects of the 
long wavelength sigma fluctuations on the fluctuations
of the pions.
We use mean field theory throughout Ref. \cite{us1}. 
The fluctuations of
the sigma field around the minimum of $\Omega(\sigma)$ are not
small; however, this does not make much difference
to the quantities of interest, all of which diverge
like $m_\sigma^{-2}\sim \xi^2$ at the critical point. 
The divergence is that of the sigma field susceptibility,
and for 
the 3d-Ising universality class we know the corresponding exponent
to be $\gamma/\nu=2-\eta$ which is $\approx2$ to 
within a few percent because $\eta$ is
small. We can therefore safely use mean-field
mean field results with their $m_\sigma^{-2}$ divergence,
and will take 
$m_\sigma\sim 1/\xi_{\rm max}\sim 1/(6 {\rm ~fm})$ in our estimates.

We now have all the ingredients in place to present our estimate of the
size of the effect of the critical fluctuations 
of the sigma field on the
fluctuations of the direct pions, via the coupling $G$.
We express the size of the effect of interest 
by rewriting the ratio $\sqrt{F}$ of (\ref{dataratio}) and (\ref{theoryratio})
as 
$$
\sqrt{F} = \sqrt{ F_B F_{\rm res} F_{\rm EC} F_\sigma}
$$
and presenting $F_\sigma$. We find:
\begin{equation}\label{FTresultmu60}
F_\sigma = 1 + 0.35 \left(G_{\rm freeze-out}\over 300\ {\rm MeV}\right)^2 
\left(\xi_{\rm freeze-out}\over 6\ {\rm fm}\right)^2 \ .
\end{equation}
where we have taken $T=120$ MeV and $\mu_\pi=60$ MeV.
$F_\sigma$ will be reduced by about a factor of two,
because 
not all of the pions which
are observed are direct. The coupling $G$ transmits the
effects of the critical $\sigma$ fluctuations to the pions at 
freezeout, not to the (heavier) resonances.
The size of the effect depends quadratically on the coupling $G$.
We argued above that $G$ is reduced
to $G_E\sim 300$ MeV at the critical point. However, freeze-out
may occur away from the critical point, in which
case $G$ would be larger, although still much
smaller than its vacuum value.  
The size of the effect also depends quadratically
on the sigma correlation length at freeze-out, and we have
seen that there are many caveats in an estimate like 
$\xi_{\rm freeze-out}\sim\xi_{\rm max}\sim 6$ fm.

We have studied two different effects of the critical
fluctuations on 
$\sqrt{F}$.  First, $F_{\rm EC}\rightarrow 1$, leading to
about a $1\%$ increase in $\sqrt{F}$.
The direct effect of
the critical fluctuations 
is a much larger increase in 
$\sqrt{F}$
by a factor of $\sqrt{F_\sigma}$. We have displayed the
various uncertainties in the factors contributing to our estimate 
(\ref{FTresultmu60}) so that when
an experimental detection of an increase 
and then subsequent decrease in 
$\sqrt{F}$
occurs, as control parameters are varied and the critical point
is approached and then passed, we will be able to use 
the measured magnitude of this nonmonotonic effect to constrain
these uncertainties.  It should already be clear that an 
effect as large as $10\%$ in $\sqrt{F_\sigma}$ is easily possible; 
this would be 50 times larger than the statistical error in
the present data. 

We now give a brief account of the effect of critical fluctuations 
on $\langle (\Delta N)^2\rangle$ and $\langle \Delta N \Delta p_T\rangle$.
The contribution of the direct pions to $\langle (\Delta N)^2\rangle$
can easily double, but the multiplicity fluctuations are 
dominated by the pions from resonance decay, so we 
estimate that the critical multiplicity fluctuations
lead to about a 10-20\% increase in $\langle (\Delta N)^2\rangle$.
(This neglects the pions from sigma decay. See below.)
The cross-correlation $\langle \Delta N \Delta p_T\rangle$
only receives contributions from nontrivial
effects, and we find  that near the critical point, the contribution
from the interaction with the sigma field is dominant.
We estimate that (for $G_{\rm freeze-out}\sim 300$ MeV and 
$\xi_{\rm freeze-out}\sim 6$ fm) the cross-correlation will 
be a factor of 10-15 times larger than in the absence
of critical fluctuations!
The lesson is clear: although
this correlation is small, it may 
increase in magnitude by a very large
factor near the critical point.

The effects of the critical fluctuations can be detected
in a number of ways. First, one can find
a nonmonotonic increase in  $F_\sigma$, the
suitably normalized increase in the variance of event-by-event
fluctuations of the mean transverse momentum.  
Second, one can find a nonmonotonic increase in 
$\langle (\Delta N)^2\rangle$. Both these effects
can easily be between one and two orders of magnitude 
greater than the statistical errors in present
data. Third, one can
find a nonmonotonic increase in the magnitude
of $\langle \Delta p_T\Delta N\rangle$.  This quantity
is small, and it has not yet been demonstrated that it
can be measured. However, it may change at the critical
point by a large factor, and is therefore worth measuring.
In addition to effects on these and many other observables,
it is perhaps most distinctive to measure the microscopic
correlator $\langle \Delta n_p \Delta n_k \rangle$ itself.
The effects proportional to $1/m_\sigma^2$ in 
has
a specific dependence on $p$ and $k$. It introduces off-diagonal correlations
in $pk$ space.  Like the off-diagonal anti-correlation 
introduced by energy conservation, this makes it easy to distinguish from the
Bose enhancement effect, which is diagonal in $pk$. 
Near the critical point, the off-diagonal anti-correlation
vanishes and the off-diagonal correlation due to sigma exchange grows.
Furthermore, the effect of $\sigma$ exchange is not restricted
to identical pions, and should be visible as correlations between
the fluctuations of $\pi^+$ and $\pi^-$.  The 
dominant diagonal term proportional to $\delta_{pk}$ 
will be absent in the correlator 
$\langle \Delta n^+_p \Delta n^-_k \rangle$, and the 
effects of
$\sigma$ exchange will be the dominant contribution to this quantity
near the critical point.

\section{Pions From Sigma Decay}

Having analyzed the effects of the sigma field on 
the fluctuations of the direct pions, we next ask what
becomes of the
sigmas themselves.  
For choices of control parameters
such that freeze-out occurs at or near the critical endpoint, 
the excitations of the sigma field, sigma (quasi)particles, 
are nearly massless at freeze-out and
are therefore numerous.  Because the pions are massive
at the critical point, these $\sigma$'s cannot immediately
decay into two pions. Instead, they persist as the
temperature and density of the system further decrease.
During the expansion, the in-medium $\sigma$ mass rises
towards its vacuum value and eventually exceeds the 
two pion threshold.  Thereafter, the $\sigma$'s decay,
yielding a population of pions which do not get a chance
to thermalize because they are produced after freeze-out.
We estimate the momentum spectrum of these pions
produced by delayed $\sigma$ decay. 
An event-by-event
analysis is not required in order to see these pions.
The excess multiplicity at low $p_T$ will appear and
then disappear in the single particle inclusive distribution
as control parameters are varied such
that the critical point is approached and then
passed.  

In calculating the inclusive single-particle $p_T$-spectrum
of the pions from sigma decay, we must treat
$m_\sigma$ as time-dependent,
and should also take $G$ to evolve with time. However,
the dominant time-dependent effect is the opening
up of the phase space for the decay as
$m_\sigma$ increases with time and 
crosses the two-pion threshold. We therefore treat
$G$ as a constant.  We have estimated that
in vacuum with  $m_\sigma=600$ MeV,  the coupling is
$G\sim 1900$ MeV, whereas at the critical end point
with $m_\sigma = 0$, the coupling is reduced, perhaps
by as much as a factor of six or so.  In this section,
we need to estimate $G$ at the time when $m_\sigma$ is
at or just above twice the pion mass.  We will use
$G\sim 1000$ MeV, recognizing that we may be off by
as much as a factor of two.  

We parametrize
the time dependence of the sigma mass by
$m_\sigma(t) = 2 m_\pi (1+ t/\tau)$,
where we have defined $t=0$ to be the time at which
$m_\sigma$ has risen to $2m_\pi$ and 
have introduced the timescale $\tau$ over which
$m_\sigma$ increases from $2m_\pi$  to $4m_\pi$.
It seems likely that $5\ {\rm fm}<\tau<20 \ {\rm fm}$.
We find that 
the mean transverse momentum of the pions produced by
sigma decay is
\begin{equation}
\langle p_T \rangle \sim 0.58 
\,m_\pi \left(\frac{1000\ {\rm MeV}}{G}\right)^{2/3}
\left(\frac{10\ {\rm fm}}{\tau}\right)^{1/3}\ .
\end{equation}
We therefore estimate that if freeze-out occurs near the critical
point, there will be a nonthermal population of pions
with transverse momenta of order half the pion mass  
with a momentum distribution given in  Ref. \cite{us1}.

How many such pions can we expect?  
This is determined by the sigma mass at freeze-out.  If 
$m_\sigma$ is comparable to $m_\pi$ at freeze-out, then there are half as many
$\sigma$'s at freeze-out as there are charged pions.  Since each
sigma decays into two pions, and two thirds of those pions
are charged, the result is that the number of charged pions
produced by sigma decays after freeze-out is $2/3$ of
the number of charged pions produced directly by the freeze-out
of the thermal pion gas.   Of course, if freeze-out occurs closer
to the critical point 
at which $m_\sigma$ can be as small as $(6 {\rm ~fm})^{-1}$, there would
be even more sigmas.  We therefore suggest that as experimenters
vary the collision energy, one way they can discover the critical
point is to see the appearance and then disappearance of a population
of 
pions with $\langle p_T\rangle \sim m_\pi/2$ which are almost as numerous
as the direct pions.
Yet again, it is the nonmonotonicity of this signature as
a function of control parameters which makes it distinctive.

The event-by-event fluctuations of the multiplicity of these
pions reflect the
fluctuations of the sigma field whence
they came \cite{SRS}.  We estimate \cite{us1} that the event-by-event
fluctuations of the multiplicity of the pions produced
in sigma decay will be 
$\langle (\Delta N)^2\rangle \approx  2.74 \langle N \rangle$.
We have already seen in that the critical fluctuations 
of the sigma field increase the fluctuations in the multiplicity
of the direct pions sufficiently that the 
increase in the fluctuation of the multiplicity of 
all the pions will be increased by about
$10-20\%$.  
We now see that in the vicinity of
the critical point, there will be a further nonmonotonic
rise in the fluctuations of the multiplicity of the 
population of pions with $\langle p_T\rangle
\sim m_\pi/2$ which are produced in
sigma decay.

\section{Outlook}

Our understanding of the thermodynamics of
QCD will be greatly enhanced by the detailed
study of event-by-event fluctuations in heavy
ion collisions.  We have estimated the influence of a number
of different physical effects, some special to the
vicinity of the critical point but many not.
The predictions of a simple resonance gas model, which
does not include critical fluctuations, are
to this point in very good agreement
with the data. More detailed study, for example with varying
cuts in addition to new observables, will help to further constrain 
the nonthermodynamic fluctuations, which are clearly small,
and better understand the different thermodynamic effects.
The signatures we analyze allow experiments
to map out distinctive features of the QCD phase diagram.
The striking example which we have considered in
detail is the effect of a second order critical end point.
The nonmonotonic
appearance and then disappearance of any one of the signatures 
of the critical fluctuations which we have described 
would be strong evidence for the critical point. 
Furthermore, 
if a nonmonotonic
variation is seen in several of these observables, then
the maxima in all the 
signatures must occur simultaneously, at the same value
of the control parameters. Simultaneous detection of the
effects of the critical fluctuations on different observables
would turn strong evidence into an unambiguous discovery.

\section*{Acknowledgments}

We are grateful to G. Roland for providing us with
preliminary NA49 data.
We acknowledge helpful conversations with 
M. Creutz, U. Heinz, M. Ga\'zdzicki, V. Koch,
St. Mr\'owczy\'nski, G. Roland and T. Trainor.

I thank the organizers of SEWM'98
for a conference which, by bringing  
together those studying QCD matter in extreme conditions
and those studying electroweak matter in extreme conditions,
was stimulating and enjoyable.

This work was supported in part by a DOE 
Outstanding Junior Investigator Award, by 
the A. P. Sloan
Foundation,  and by the DOE
under cooperative research agreement DE-FC02-94ER40818.

\end{document}